\documentclass[conference]{IEEEtran}
\IEEEoverridecommandlockouts
\usepackage{cite}
\usepackage{amsmath,amssymb,amsfonts}
\usepackage{algorithmic}
\usepackage{graphicx}
\usepackage{textcomp}
\usepackage{xcolor} 
\usepackage{multirow}
\usepackage{threeparttable}

\def\BibTeX{{\rm B\kern-.05em{\sc i\kern-.025em b}\kern-.08em
    T\kern-.1667em\lower.7ex\hbox{E}\kern-.125emX}}

\makeatletter
\let\NAT@parse\undefined
\makeatother
\usepackage{hyperref}

\begin{document}

\title{Magnitude-Phase Dual-Path Speech Enhancement Network based on Self-Supervised Embedding and Perceptual Contrast Stretch Boosting}

\author{
\IEEEauthorblockA{Alimjan Mattursun$^1$, Liejun Wang$^1$$^{\dagger}$\thanks{$^{\dagger}$Both Liejun Wang and Yinfeng Yu are corresponding authors.}, Yinfeng Yu$^1$$^{\dagger}$, Chunyang Ma$^2$}
\\
\IEEEauthorblockA{$^1$School of Computer Science and Technology, Xinjiang University, China \\ 
$^2$Cadre Division Organization Department of the Party Committee, Xinjiang Uygur Autonomous Region, China}
\IEEEauthorblockA{e-mail: alim@stu.xju.edu.cn, wljxju@xju.edu.cn, yuyinfeng@xju.edu.cn}
\thanks{This study was funded by the Excellence Program Project of Tianshan, Xinjiang Uygur Autonomous Region, China (grant number 2022TSYCLJ0036), the Central Government Guides Local Science and Technology Development Fund Projects (grant number ZYYD2022C19), and the National Natural Science Foundation of China (grant numbers 62463029, 62472368 and 62303259).}


}

\maketitle

\begin{abstract}
Speech self-supervised learning (SSL) has made great progress in various speech processing tasks, but there is still room for improvement in speech enhancement (SE). This paper presents BSP-MPNet, a dual-path framework that combines self-supervised features with magnitude-phase information for SE. The approach starts by applying the perceptual contrast stretching (PCS) algorithm to enhance the magnitude-phase spectrum. A magnitude-phase 2D coarse (MP-2DC) encoder then extracts coarse features from the enhanced spectrum. Next, a feature-separating self-supervised learning (FS-SSL) model generates self-supervised embeddings for the magnitude and phase components separately. These embeddings are fused to create cross-domain feature representations. Finally, two parallel RNN-enhanced multi-attention (REMA) mask decoders refine the features, apply them to the mask, and reconstruct the speech signal.We evaluate BSP-MPNet on the VoiceBank+DEMAND and WHAMR! datasets. Experimental results show that BSP-MPNet outperforms existing methods under various noise conditions, providing new directions for self-supervised speech enhancement research. The implementation of the BSP-MPNet code is available online\footnote[2]{https://github.com/AlimMat/BSP-MPNet. \label{s1}}
\end{abstract}


\section{INTRODUCTION}
\label{sec:intro}

In real-world acoustic environments, various forms of background noise and room reverberation significantly affect the clarity and intelligibility of speech. These issues are particularly prominent in speech-related applications such as conference systems, speech recognition, speaker identification, and hearing aids\cite{review1}. Therefore, improving the quality and intelligibility of noisy speech has become an important research topic in speech processing. The goal of speech enhancement (SE) is to extract clean speech from noisy signals, thereby improving the overall speech quality\cite{review1}. With the rapid development of deep learning, deep learning-based speech enhancement methods have become a research hotspot.

\begin{figure}[htbp]
    \begin{minipage}[b]{1.0\linewidth}
    \centering
    \centerline{\includegraphics[width=9cm,height=3.3cm]{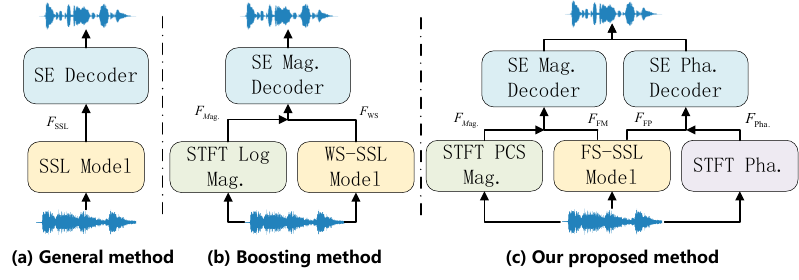}}
    \end{minipage}
    \caption{The speech enhancement network based on self-supervised embedding. (a) is a general method. (b) is the boosting method added to the magnitude spectrum. (c) is our proposed magnitude-phase dual-path enhancement method.}
    \label{fig:f1}
	
\end{figure}

Deep learning-based speech enhancement methods can be categorized into three types: time-domain methods\cite{WAVCRN,SECONFORMER,MANNER}, time-frequency (T-F) domain methods\cite{METRICGAN+,CMGAN,HATFANET}, and cross-domain methods\cite{BSSSE,BSSCFFMA,BSSCSPCS}. Time-domain methods directly estimate clean speech from noisy speech waveforms, enabling end-to-end modeling. Time-frequency domain methods convert speech signals to the frequency domain using short-time Fourier transform (STFT), enhance the speech in the spectrogram, and reconstruct the speech signal through inverse transformation. Cross-domain methods combine features from different domains, leveraging acoustic and noise information across domains to improve enhancement performance and robustness to noise.

Despite significant progress with time-domain and time-frequency methods in speech enhancement, these approaches typically rely on large amounts of labeled data for supervised learning. However, obtaining large-scale labeled speech data is expensive and labor-intensive. To address this, self-supervised learning (SSL) has gradually become an effective alternative. SSL extracts useful feature representations from unlabeled datasets, reducing the dependency on labeled data, and has been widely applied in speech recognition, emotion recognition, and other tasks. Common self-supervised speech models include Wav2vec2.0\cite{WAV2VEC20}, WavLM\cite{WAVLM}, and multimodal models like Data2vec\cite{DATA2VEC} and Data2vec2.0\cite{DATA2VEC2.0}. As shown in Fig. \ref{fig:f1}, self-supervised learning models are increasingly utilized in SE tasks. However, most studies primarily focus on magnitude spectrum enhancement, often overlooking phase spectrum processing\cite{BSSSE,BSSCFFMA}.
Research has shown that the phase spectrum plays a crucial role in reconstructing high-quality speech, and optimizing the magnitude spectrum alone can lead to phase distortion, negatively affecting speech quality\cite{MPSENET}. Therefore, integrating phase spectrum processing within an SSL framework for dual-path enhancement remains an important area of research. 
To this end, this paper proposes BSP-MPNet.


The main contributions of this work are as follows: 
\begin{itemize}
	\item \textbf{MP-2DC Encoder:} coarse-grained feature extraction of the \textbf{PCS-boosted} magnitude-phase spectra, addressing the issue of missing fine-grained information in SSL models and alleviating feature mismatch and representation distortion during cross-domain feature fusion.
	
	\item \textbf{FS-SSL:} decouples self-supervised features into magnitude- and phase-related representations, combined with the magnitude-phase spectra extracted by MP-2DC to provide richer cross-domain features for the decoder.
	
	\item \textbf{REMA Mask Decoder:} integrates RNN, self-attention, and time-frequency attention to process magnitude and phase features separately. Its parallel masking mechanism captures global and local dependencies, improving speech reconstruction in noisy conditions. The effectiveness of BSP-MPNet is validated on two datasets of varying complexity.
\end{itemize}
\begin{figure*}[htbp]
	\begin{minipage}[b]{1.0\linewidth}
		\centering
		\centerline{\includegraphics[width=18cm]{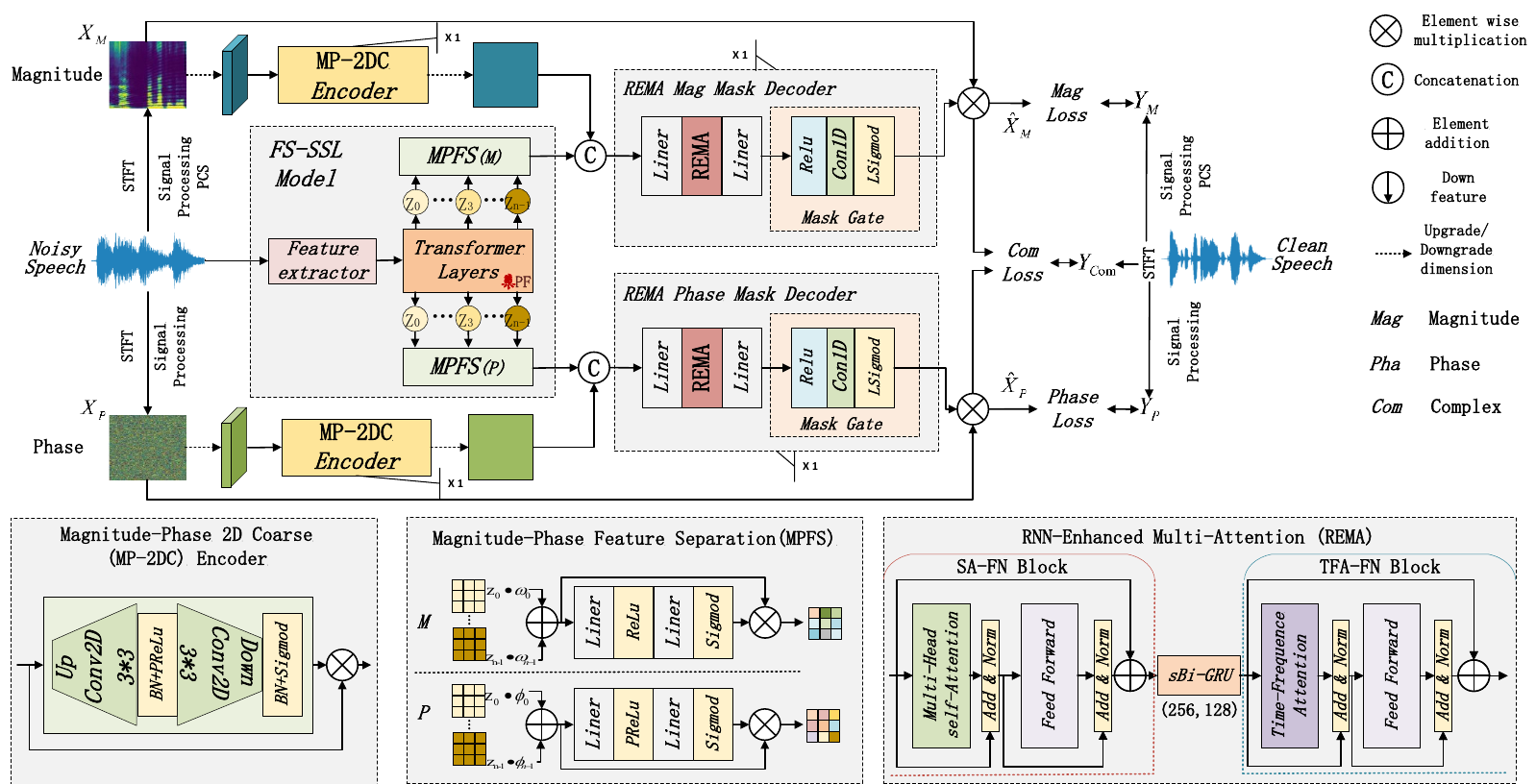}}
		\caption{Architecture of the proposed BSP-MPNet. "STFT" represents the short-time fourier transform of the speech.}\medskip
		\label{fig:f2}
	\end{minipage}
\end{figure*}

\section{RELATED WORK}
\label{sec:rework}

\subsection{SSL Model and Fine-Tuning}

Self-supervised learning (SSL) improves performance in downstream speech tasks through fine-tuning techniques, such as speech recognition and enhancement. In this study, we select the discriminative model WavLM\cite{WAVLM} and the multi-task learning model Data2Vec\cite{DATA2VEC} as the base SSL models to extract latent speech representations and use the partially fine-tuned (PF) approach to improve further the SSL model's performance in speech enhancement tasks.

\subsection{PCS and Cross-Domain Features}

Perceptual Contrast Stretching (PCS)\cite{PCS} is a spectral processing technique based on the frequency sensitivity of the human auditory system. It enhances the contrast of the spectrum to highlight key details in the speech signal, thereby improving its perceptibility and clarity. as follows:
\begin{equation}
	X^{'}(\omega,t)= log(|STFT(x)|+1) ,\label{eq1}
\end{equation}
\begin{equation}
	\hat{X}(\omega,t)= X^{'}(\omega,t) \cdot W(\omega,t) ,\label{eq2}
\end{equation}
where $x$ is the speech waveform, $X^{'}(\omega,t)$ is the compressed magnitude spectrum, $W(\omega,t)$ is the perceptual score from the Band Importance Function (BIF)\cite{PCS}, and $\hat{X}(\omega,t)$ is the result of PCS processing.

Cross-domain features significantly improve the performance of automatic speech recognition (ASR) and SE compared to single-domain speech features\cite{BSSSE}. Therefore, this paper combines FS-SSL features with PCS-optimized magnitude-phase spectrum features to enrich the speech and noise information, further improving the model's performance in SE tasks.

\section{PROPOSED METHOD}
\label{sec:prometh}
The proposed BSP-MPNet method is illustrated in Fig. \ref{fig:f2}. It adopts two parallel encoder-decoder architectures to denoise both magnitude and phase simultaneously. First, the MP-2D encoder performs global coarse-grained feature extraction of the PCS-boosted magnitude-phase spectra in the time-frequency direction. The FS-SSL model then extracts self-supervised latent features related to the magnitude-phase. These features are subsequently fused to obtain cross-domain magnitude-phase features. Finally, the REMA mask decoder extracts both global and local features of the cross-domain representation in the time-frequency direction, captures temporal dependencies, and applies a mask with the original spectrogram. The enhanced magnitude and phase are reconstructed to recover the speech signal.

\subsection{Magnitude-Phase 2D Coarse (MP-2DC) Encoder}

The effectiveness of SSL in SE tasks has been well established\cite{SSLSESP}. By incorporating magnitude spectra, SSL addresses the issue of lacking fine-grained information in its features\cite{BSSSE}. 
However, compatibility issues between cross-domain features may lead to feature distortion and degrade SE performance.
Moreover, deconvolution has proven to be highly effective in SE tasks, as it recovers spatial details and enhances the resolution of time-frequency features\cite{IFFNET}. Based on these findings, we introduce MP-2DC encoder to address these issues further.

As shown in Fig. \ref{fig:f2}, the MP-2DC encoder consists of dimensionality-increasing 2D convolutions (Up Conv2D), dimensionality-reducing 2D convolutions (Down Conv2D), batch-normalization layers, and activation functions. 
Specifically, first, the dimensions of the magnitude and phase spectra are expanded, changing the original input shape $X_m \sim X_p \in \mathbb{R} ^{B*C*T} \longrightarrow  X_{m}^{'} \sim X_{p}^{'} \in \mathbb{R} ^{B*1*F*T}$. Next, a dimensionality-increasing 2D convolution is used for feature extraction, resulting in a feature map of size $X_{m}^{'} \sim X_{p}^{'} \in \mathbb{R} ^{B*16*F*T}$. Then, the feature is normalized and activated using PReLu. Finally, dimensionality-reducing 2D convolutions refine the features, restoring them to the $X_{m}^{'} \sim X_{p}^{'} \in \mathbb{R} ^{B*1*F*T}$ dimension and a Sigmoid activation function are applied to multiply the features with the original input and obtain magnitude-phase correlated latent features. To enable the fusion of SSL features, the latent features are further reduced to a dimension of $X_{m}^{'} \sim X_{p}^{'} \in \mathbb{R}^{B*1*F*T} \longrightarrow X_{m}^{'} \sim X_{p}^{'}\in \mathbb{R} ^{B*C*T}$
	
\subsection{Feature-Separated SSL (FS-SSL) Model}
The weighted-sum-based SSL model shows significant advantages in downstream SE tasks. Additionally, self-supervised features have a similar distribution to traditional Fbank features\cite{BSSSE}. Other studies have also verified the high reliability of the fusion of self-supervised features and magnitude spectrum features\cite{BSSCFFMA}. However, no research has explored the relationship between self-supervised features and phase information. This raises a key question: do self-supervised features contain or can extract phase-related representations? This paper experimentally verifies this and concludes that self-supervised features can carry phase information or can, to some extent, capture phase-related feature representations.
	
Based on these findings, we propose an FS-SSL model and design an magnitude-phase feature separation (MPFS) module, as shown in Fig. \ref{fig:f2}. The model consists of two core parts: first, the transformer layers are filtered layer in the SSL model by layer to select and weigh the feature outputs from each layer adaptively, as follows:
\begin{equation}
	F_{m} = \sum_{i=0}^{N-1} \left[ \omega_i*z_i \right],\;\;\;\;
	F_{p} = \sum_{i=0}^{N-1} \left[ \phi_i*z_i \right],\label{eq3}
\end{equation}
where $F_{m} \in \mathbb{R} ^{B*D*T}$ is magnitude-related features,  $F_{p} \in \mathbb{R} ^{B*D*T}$ is phase-related features. $i$ is the number of Transformer layers in SSL. $z_i$ each Transformer layer's output features in SSL. $\omega_i$ and $\phi_i$ are the learnable weights of the corresponding parameters $0 \le \omega_i \sim \phi_i \le 1, \sum_{i} \omega_i \sim \sum_{i}\phi_i = 1 $.
	
Next, fine-grained filtering is applied to the generated overall feature representations to optimize each feature value further. This feature-separation modeling strategy helps enhance the adaptability of self-supervised features in both magnitude and phase representations, meeting the need for fine-grained features in speech enhancement tasks, as follows:
\begin{equation}
F_{m}'=Sigmod(Liner(ReLu(Liner(F_{m})))) \odot F_{m},\label{eq4}
\end{equation}
\begin{equation}
F_{p}'= Sigmod(Liner(ReLu(Liner(F_{p})))) \odot F_{p},\label{eq5}
\end{equation}	
where $F_{m}^{'} \in \mathbb{R} ^{B*D*T}$ is magnitude-related latent features, $F_{p}^{'} \in \mathbb{R} ^{B*D*T}$ is phase-related latent features.
	
\subsection{RNN-Enhanced Multi-Attention (REMA) Mask Decoder}
	
RNNs have been widely applied to time-series tasks\cite{SAAVN}. 
Additionally, SE methods based on Transformer architectures have demonstrated exceptional performance\cite{CMGAN}, with the self-attention mechanism playing a crucial role. However, multiple attention mechanisms, compared to a single one, can more effectively capture features and significantly improve decoding performance\cite{MANNER}. Therefore, we propose a REMA mask decoder to extract more amplitude-phase mask information from cross-domain features and enhance decoding performance. 
	
The structure of the REMA module is shown in Fig.\ref{fig:f2}. It is based on an improved version of the Squeezeformer\cite{SQUEEZEFORMER} architecture. It consists of several core submodules: the multi-head self-attention (MHSA) module, the feed-forward network (FFN) module, a small bidirectional GRU (sBi-GRU), the time-frequency attention (TFA) module, and the mask gate. We combine MHSA, TFA, and FFN to enhance flexibility in constructing the SA-FN and TFA-FN blocks. Each module is normalized using post-layer normalization, with multiple residual connections employed to optimize the model structure, thereby accelerating convergence and improving stability.

First, the magnitude-phase related latent features obtained from FS-SSL and MP-2DC are concatenated to form the cross-domain feature $Z_{cross} \in \mathbb{R} ^{B*(D+C)*T}$, which is then fed into the SA-FN Block to produce $Z_{sa}$.
\begin{equation}
    Z_{sa}=SA\text{-}FN(Z_{cross}),\label{eq6}
\end{equation}
Next, $Z_{sa}$ is input into a sBi-GRU to extract sequential information, resulting in $Z_{gru}$.
\begin{equation}
    Z_{gru}=sBi\text{-}GRU(Z_{sa}),\label{eq7}
\end{equation}
Subsequently, $Z_{gru}$is fed into the TFA-FN Block for time-frequency (channel) processing, producing $Z_{tfa}$.
\begin{equation}
	Z_{tfa}=TFA\text{-}FN(Z_{gru}),\label{eq8}
\end{equation}
Finally, $Z_{tfa}$ is through a linear layer to obtain the REMA output $Z_{rema}$, which is then fed into the masking gate to compute the mask feature $Z_{mask}$. The mask feature is element-wise multiplied with the original magnitude-phase spectrum to yield the enhanced magnitude-phase spectrum.
\subsubsection{Time-Frequency Attention (TFA) Module}
\begin{figure}[htbp]
	\begin{minipage}[b]{1.0\linewidth}
		\centering
        \centerline{\includegraphics[width=8.5cm,height=5.5cm]{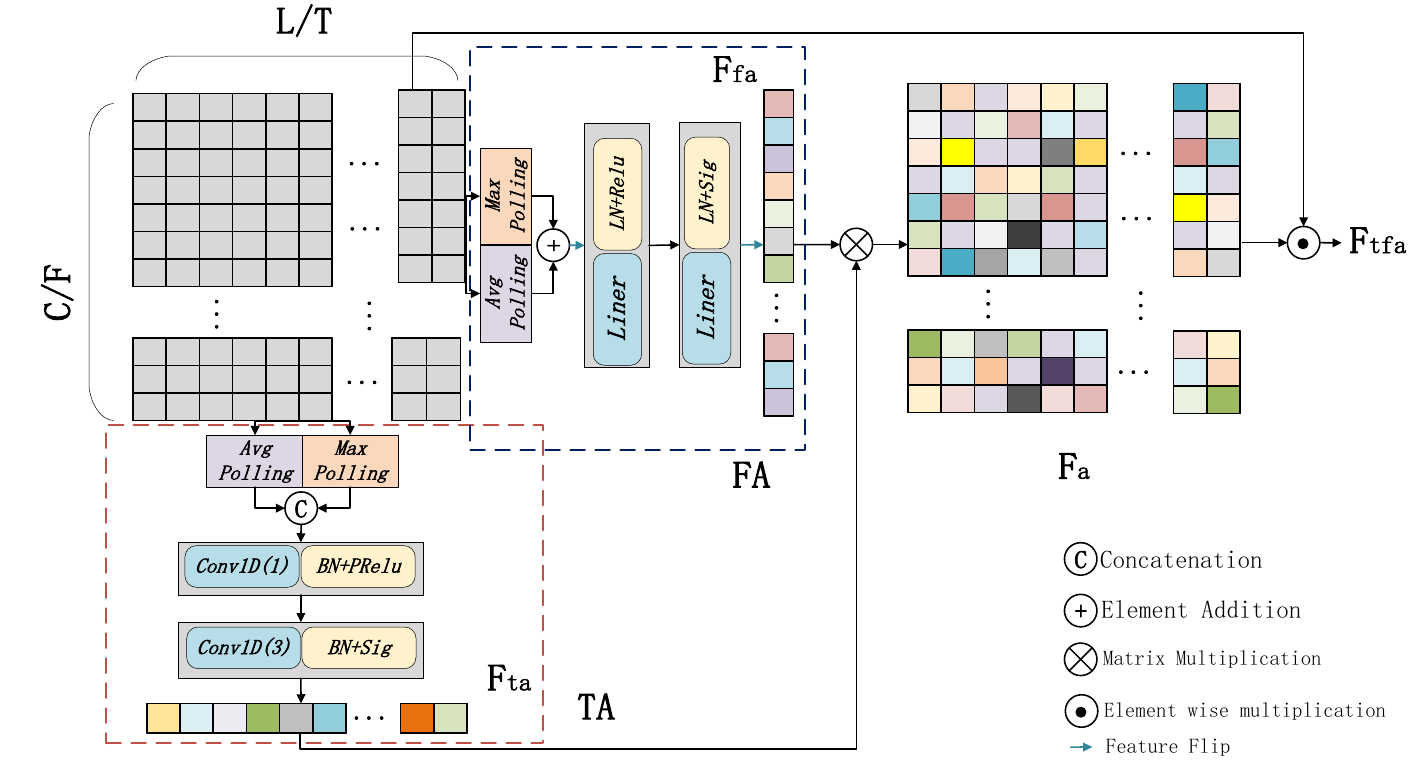}}
	\end{minipage}
	\caption{Structure of Time-Frequency Attention (TFA) Module.}
		\label{fig:f3}
\end{figure}

As shown in Fig. \ref{fig:f3}, the TFA module consists of two components: Time Attention (TA) and Frequency Attention (FA).
In the TA module, the input feature $F_{in} \in \mathbb{R} ^{B*F*T}$ is first compressed along the frequency (channel) dimension by applying max pooling and average pooling, and their results are concatenated to obtain the compressed feature $F_{concet} \in \mathbb{R} ^{B*2*T}$. 
$F_{concet}$ is then passed through two convolutional layers and activation functions to obtain the time attention weights $F_{fa} \in \mathbb{R} ^{B*1*T}$.
In the FA module, the feature  $F_{in} \in \mathbb{R} ^{B*F*T}$ is compressed along the time dimension by applying max pooling and average pooling, and their results are summed to obtain the compressed feature $F_{add} \in \mathbb{R} ^{B*F*1}$ $\longrightarrow$ $F_{add} \in \mathbb{R} ^{B*1*F}$. $F_{add}$ is then passed through two liner layers and activation functions to obtain the frequency (channel) attention weights $F_{fa} \in \mathbb{R} ^{B*1*F}$ $\longrightarrow$ $F_{fa} \in \mathbb{R} ^{B*F*1}$.
Finally, $F_{ta}$ and $F_{fa}$ are multiplied to obtain the expanded time-frequency attention weights $F_{a} \in \mathbb{R} ^{B*F*T}$, which are then element-wise multiplied with the original input feature $F_{in}$ to produce the final time-frequency attention representation $F_{tfa} \in \mathbb{R} ^{B*F*T} $.
	
\subsubsection{Mask Gate}
	
Most deep learning-based mask-based SE methods use activation functions such as ReLU and sigmoid to generate mask features, which yields good results\cite{CMGAN,HATFANET,BSSSE,BSSCFFMA}. However, recent studies have shown that the LSigmoid function performs better in mask generation\cite{MPSENET}. Based on these findings, we designed a simplified mask generation module to compute masks for both magnitude and phase accurately. The module consists of a ReLU activation function, a 1D convolutional layer, and an LSigmoid\cite{MPSENET} function, as detailed below:
\begin{equation}
	Z_{mask}=LSigmod(Conv1D(ReLu(Z_{rema}))),\label{eq9}
\end{equation}
where $Z_{mask}$ is the output of the mask gate, $Z_{rema}$ is the output of the REMA module, and the LSigmoid function is given by the following formula:
\begin{equation}
	LSigmod(t) = \dfrac{\beta}{1+e^{1-\alpha t}} ,\label{eq10}
\end{equation}	
where $\beta$ is set to 1.0, and $\alpha \in \mathbb{R} ^{F}$ is a trainable parameter, allowing the model to adaptively modify the shape of the activation function within different frequency bands.

\subsection{Loss Function}
	
We defined a multi-level loss function to train the proposed BSP-MPNet. To achieve simultaneous optimization at the magnitude, phase, and complex spectrum levels, we used magnitude loss $\mathcal{L}_{Mag.}$, phase loss $\mathcal{L}_{Pha.}$, and complex spectrum loss $\mathcal{L}_{Com.}$, as follows:
\begin{equation}
	\mathcal{L}_{Mag.} = \mathbb{E}_{y_m,\hat{x}_m}[ ||y_{m}-\hat{x}_{m} ||_1 ],\label{eq11}
\end{equation}
\vspace{0.01cm}
\begin{equation}
	\mathcal{L}_{Pha.} = \mathbb{E}_{y_p,\hat{x}_p}[ ||\mathcal{K}_{AW}(y_{p}-\hat{x}_{p}) ||_1 ],\label{eq12}
\end{equation}
\vspace{0.01cm}
\begin{equation}
	\mathcal{L}_{Com.} = \mathbb{E}_{y_r,\hat{x}_r}[ ||y_{r}-\hat{x}_{r} ||^2_2] + \mathbb{E}_{y_i,\hat{x}_i}[||y_{i}-\hat{x}_{i} ||^2_2],\label{eq13}
\end{equation}
where $y$ is the clean speech signal, $\hat{x}$ is the enhanced speech signal, and $\mathcal{K}_{AW}(t)$ \cite{MPSENET}is an anti-wrapping function used to prevent errors caused by phase wrapping, defined as:
	
\begin{equation}
	\mathcal{K}_{AW}(t) = |t-2\pi*round(\frac{t}{2\pi})|,\label{eq14}
\end{equation}
where $t \in \mathbb{R} $.
	
The final loss function is a linear combination of the above losses, which is:
\begin{equation}
	\mathcal{L}_{loss} = \lambda_1\mathcal{L}_{Mag.}+\lambda_2\mathcal{L}_{Pha.}+\lambda_3\mathcal{L}_{Com.},\label{eq15}
\end{equation}
where $\lambda_1,\lambda_2,\lambda_3$ are hyperparameters set to 1, 0.5, and 0.1, respectively, in the empirical experiments.
\section{EXPERIMENTS}

\subsection{Dataset}

\begin{table}[htbp]
\centering
\renewcommand\arraystretch{1.3}
\caption{Comparison results on the VoiceBank+DEMAND dataset regarding objective speech quality metrics. The SSL models utilize the Data2vec and WavLM (with fine-tuning).}
\label{tab:t1}
\resizebox{8.5cm}{!}{
    \begin{tabular}{lcccccc}
	\hline
	\textbf{Methods} & \textbf{Domain} & \textbf{PESQ}$\uparrow$ & \textbf{CSIG}$\uparrow$ & \textbf{CBAK}$\uparrow$ & \textbf{COVL}$\uparrow$ & \textbf{STOI}(\%)$\uparrow$ 
    \\
    \hline
    Noisy  & - & 1.91 & 3.35 &	2.44 &	2.63	& 91.5 \\
	\hline
    SRTNET\cite{WAVCRN}  &  & 2.69 &	4.12 &	3.19 &	3.39	& - \\
    SE-Conformer\cite{SECONFORMER}  & Time  & 3.12  & 4.18  & 3.47  & 3.51 & 95.0 \\
    MNNER\cite{MANNER}  &  & 3.21  &	4.53  &	3.65  &	3.91 & 95.0 \\
    \cline{2-7}
    MetricGAN+\cite{METRICGAN+} &   & 3.15  & 4.14  & 3.16  & 3.64  & - \\
    HATFANET\cite{HATFANET} & T-F  & 3.20 &	4.53 & 3.60 &	3.88 &	95.4 \\
    CMGAN\cite{CMGAN} &   & 3.50  & 4.63  & 3.94  & 4.12  & 96.0 \\
    \cline{2-7}
    SSL-SESP(L)\cite{SSLSESP} &   & 2.80 & - & - & - & - \\
    BSS-SE(L)\cite{BSSSE}  & Cross  & 3.20 &	4.53 &	3.60	& 3.88 & 95.7 \\
    BSS-CFFMA(B)\cite{BSSCFFMA}  &   & 3.21  & 4.55 & 3.70  & 3.91 & 94.8 \\
    BSS-CS-PCS(L)\cite{BSSCSPCS} &   & 3.54 &	\textbf{4.75} &	3.54	& \textbf{4.25} &	96.0 \\
    \hline
    \textbf{BSP-MPNet} &   &  &   &   &   &  \\
    \;\;\textbf{+Data2vec} & Cross  & 3.51  & 4.57 & 3.88  & 3.96  &  97.8\\
    \;\;\textbf{+WavLM(B)} &   & 3.61 & 4.61  & 3.91  & 4.13  & 98.2 \\
    \;\;\textbf{+WavLM(L)} &   & \textbf{3.65} & 4.66  & \textbf{3.95}  & 4.18  & \textbf{98.5} \\
    \hline
    \end{tabular}
    }
    \begin{threeparttable}
        \begin{tablenotes}
            \footnotesize\tiny
            \item L indicates the large model, while B indicates the base model.
        \end{tablenotes}
    \end{threeparttable}
\end{table}

We evaluated the enhancement performance of BSP-MPNet on the VoiceBank+DEMAND\cite{VB} and WHAMR!\cite{WHAMR} datasets. The former contains 11,572 speech segments, with segments from 28 speakers used for training and segments from 2 speakers used for testing. In the training phase, 10 types of noise are mixed with clean speech at signal-to-noise ratios (SNR) ranging from [0, 15] dB. In the testing phase, 5 types of noise are mixed at SNR ranging from [2.5, 17.5] dB. The WHAMR! dataset is an extended version of the WSJ0-2mix\cite{wsj0} dataset, containing combinations of \textbf{Noise}, \textbf{Reverberation}, and \textbf{Noise+Reverberation}. 
The SNR ranges from [-6, 3] dB, with speech, noise, and reverberation randomly mixed. The WHAMR! dataset consists of a training set with 20,000 speech segments, a validation set with 5,000 segments, and a test set with 3,000 segments.

\subsection{Evaluation Metrics}
To evaluate the performance of BSP-MPNet, we used multiple evaluation metrics. For the traditional evaluation, we adopted Perceptual Evaluation of Speech Quality (PESQ)\cite{PESQ}, Scale-Invariant Signal-to-Noise Ratio (SI-SNR)\cite{SISNR}, Short-Time Objective Intelligibility (STOI)\cite{STOI}, Speech Signal Distortion (CSIG)\cite{ALL}, Background Intrusiveness (CBAK)\cite{ALL}, and Overall Quality (COVL)\cite{ALL}. For the DNSMOS evaluation, we adopted DNSMOS P.835\cite{DNSMOS}. 
Additionally, we also introduced LLR, Cosine Similarity, and RTF.

\subsection{Experimental Setup}
All speech signals were downsampled to 16 kHz. During the training phase, 2-second audio segments were randomly selected from each speech sample, with 100 training epochs. during testing, the original length of the speech signals was used. The parameters for STFT and ISTFT were set as follows: FFT length of 32 ms, window length of 25 ms, and hop size of 6.25 ms. The batch size was set to 16. We used the Adam optimizer with a dynamic learning rate strategy\cite{IFFNET}. The training was conducted on two NVIDIA T4 GPUs, with each epoch taking approximately 7 minutes.


\begin{table}[htbp]
\centering
\renewcommand\arraystretch{1.3}
\caption{Comparison results on the VoiceBank+DEMAND dataset regarding DNSMOS and Similarity metrics. The SSL models utilize the Data2vec and WavLM (with fine-tuning).}
\label{tab:t3}
\resizebox{8.5cm}{!}{
\begin{tabular}{l|c|cccc|cc}
    \hline
    \textbf{Methods} & \textbf{Type} & \multicolumn{4}{c|}{\textbf{DNSMOS}} & \multicolumn{2}{c}{\textbf{Similarity}} \\
    \cline{3-8}
    & & \textbf{OVRL}$\uparrow$ & \textbf{SIG}$\uparrow$ & \textbf{BAK}$\uparrow$ & \textbf{P808.MOS}$\uparrow$ & 
    \textbf{LLR}$\downarrow$ & \textbf{Cosine}$\uparrow$ \\
    \hline
    \multirow{2}{*}{Ground Truth} & Clean  & 3.23 & 3.52 & 4.03 & 3.50 & 0.0 & 1.0 \\
                                   & Noisy  & 2.59 & 3.22 & 3.05 & 2.96 & 0.144 & 0.911 \\
    \hline
    \multirow{2}{*}{\textbf{BSP-MPNet}} 
    & Data2vec & 2.99  & 3.27 & 3.98  & 3.31 & 0.019 & 0.984 \\
    & WavLM(L) & 3.05  & 3.32 & \textbf{4.00}  & 3.37 & \textbf{0.015} & \textbf{0.995} \\
    \hline
\end{tabular}
}
\end{table}

\begin{table}[htbp]
\centering
\renewcommand\arraystretch{1.5}
\caption{Comparison results on the WHAMR! dataset in terms of objective speech quality metrics. The SSL models utilize the Large WavLM (with fine-tuning).}
\label{tab:t2}
\resizebox{8.5cm}{!}{
    \begin{tabular}{l|ccc|ccc|ccc}
    \hline 
    \multicolumn{1}{c} {} & \multicolumn{3}{c}{\textbf{Reverb}} & \multicolumn{3}{c}{\textbf{Noisy}} & \multicolumn{3}{c}{\textbf{Reverb+Noisy}}  \\
    \hline
    \textbf{Methods} & \textbf{PESQ}$\uparrow$  & \textbf{STOI}(\%)$\uparrow$ & \textbf{SI-SNR}$\uparrow$  & \textbf{PESQ}$\uparrow$  & \textbf{STOI}(\%)$\uparrow$ & \textbf{SI-SNR}$\uparrow$ & \textbf{PESQ}$\uparrow$  & \textbf{STOI}(\%)$\uparrow$ & \textbf{SI-SNR}$\uparrow$ \\
    \hline
    Mixed & 2.16 &	91 &	4.38 &	1.11 &	76 &	-0.99 &	1.11 &	73 &	-2.73 \\
    \hline
    DCCRN\cite{IFFNET} & 2.55 &	95 & \textbf{7.51} &	1.66 &	90 &	9.03 &	1.59 &	88 & \textbf{5.20} \\
    TSTNN\cite{IFFNET} & 2.66 &	95 &	3.56 &	1.94 &	93 &	4.17 &	1.91 &	91 &	2.89 \\
    BSS-SE\cite{BSSCFFMA} &3.02  & 91 & 5.90 & 1.84 & 89 & 7.52  &1.70  & 86 &2.16  \\
    BSS-CFFMA\cite{BSSCFFMA} &3.26  & 96 & 6.24 & 2.05 & 92 & 9.30  &1.92  & 91 &3.55  \\
    \hline
    \textbf{BSP-MPNet} & &  &  & &  & &  & & \\
    \;\;\textbf{+WavLM(L)} & \textbf{3.31}	& \textbf{96} & 6.01 &\textbf{2.30} & \textbf{93} &\textbf{11.17} & \textbf{2.05} &\textbf{91} &4.03\\
			\hline
		\end{tabular}
	}
\end{table}

\section{RESULTS AND COMPARISON}
\subsection{Performance Comparison on Two Datasets}
We first compared the denoising performance of the proposed BSP-MPNet with ten methods from three domains using traditional metrics on the Voicebank+DEMAND dataset. As shown in Table \ref{tab:t1}, BSP-MPNet outperforms other baseline methods in most metrics. Compared to the BSS-CS-PCS\cite{BSSCSPCS} method, which also uses PCS, our method improves PESQ by 0.11 points. Additionally, BSP-MPNet significantly outperforms other methods in CBAK and STOI, indicating its advantage in background noise suppression and speech intelligibility. The performance in CSIG is comparable to that of the typical CMGAN\cite{CMGAN} method.

We also used DNSMOS and similarity for evaluation, as shown in Table \ref{tab:t3}. The values were calculated for clean, noisy, and enhanced speech. The results show that our method achieves performance close to clean speech in the BAK and similarity metric and the P808.MOS value is also promising.

Finally, we evaluated the performance of BSP-MPNet's denoising, dereverberation, and joint denoising and dereverberation on the WHAMR! dataset. As shown in Table \ref{tab:t2}, BSP-MPNet outperforms other baselines on most indicators in three testing scenarios: \textbf{Noise}, \textbf{Reverb}, and \textbf{Noise+Reverb}. Further validating its enhancement performance in complex environments, and found reverberation more challenging.

\subsection{Ablation Study}
To evaluate the contribution of each key module in BSP-MPNet, we performed ablation studies, with the results presented in Table \ref{tab:t4}. When the magnitude spectrum was not enhanced (w/o Mag.), the performance metrics dropped significantly. Similarly, the absence of PCS (w/o PCS) led to a notable decline, though the performance remained higher than other methods without PCS, demonstrating the model's superiority. Furthermore, excluding phase spectrum enhancement (w/o Pha.) caused a significant performance drop, highlighting the importance of phase enhancement in speech enhancement tasks. 
\begin{table}[htbp]
	\centering
	\renewcommand\arraystretch{1.3}
	\caption{Ablation study on the  VoiceBank+DEMAND dataset. The SSL models utilize the Large WavLM.}
	\label{tab:t4}
	\resizebox{8.5cm}{!}{
	\begin{tabular}{lcccccc}
		\hline
		\textbf{Methods} & \textbf{PESQ}$\uparrow$ &	\textbf{CSIG}$\uparrow$ & \textbf{CBAK}$\uparrow$	& \textbf{COVL}$\uparrow$ & \textbf{SI-SNR}$\uparrow$ &
        \textbf{RTF}$\downarrow$
        \\
	\hline
		Noisy & 1.91 &3.35 &	2.44 &	2.63 & - &- \\
	\hline
		BSP-MPNet & 3.65 &	4.66 &	3.95 & 4.18 & 24.22 & 0.0455 \\
		\hline
		\;\;w/o PF & 3.57 &4.61	 &3.91	 &4.13	 &24.12  & 0.0455 \\
		\;\;w/o PCS & 3.31 &4.56	 &3.73	 &3.96	 &19.48 & 0.0395 \\
		\;\;w/o FS-SSL & 3.52 & 4.56 & 3.77 & 4.08 & 24.05  & 0.0350 \\
		\;\;w/o MP-2DC & 3.58 & 4.61 & 3.91 & 4.12 & 24.15  & 0.0345\\
        \;\;w/o Mag. & 2.80 &4.13 &3.17 &3.47 & 17.75 & 0.0275 \\
		\;\;w/o Pha. & 3.39 & 4.53 & 3.80 & 4.03 & 23.61 & 0.0289 \\
	\;\;w/o REMA & 3.45 & 4.48 &  3.70 & 4.01 & 23.65  & 0.0243\\
		\;\;w/o Mask Gate & 3.60 & 4.63 & 3.91 & 4.15 & 24.18 & 0.0425\\
		\hline
	\end{tabular}
		}
\end{table}
Other ablation experiments showed varying degrees of performance decline when individual modules were removed, indicating that the collaboration of all modules enables the model to achieve state-of-the-art performance.

\begin{figure}[t]
	\begin{minipage}[b]{.48\linewidth}
		\centering
		\centerline{\includegraphics[width=4.0cm]{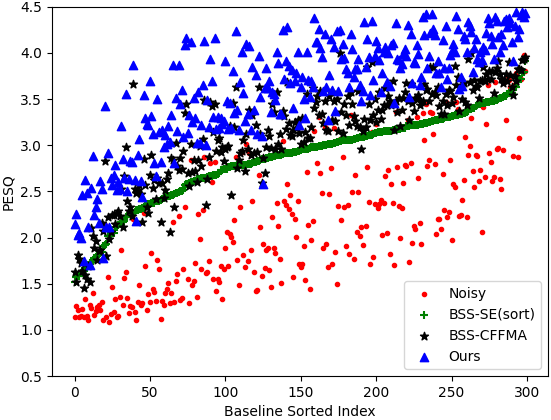}}
		\centerline{(a) Comparison of PESQ}\medskip
	\end{minipage}
	\hfill
	\begin{minipage}[b]{0.48\linewidth}
		\centering
		\centerline{\includegraphics[width=4.0cm]{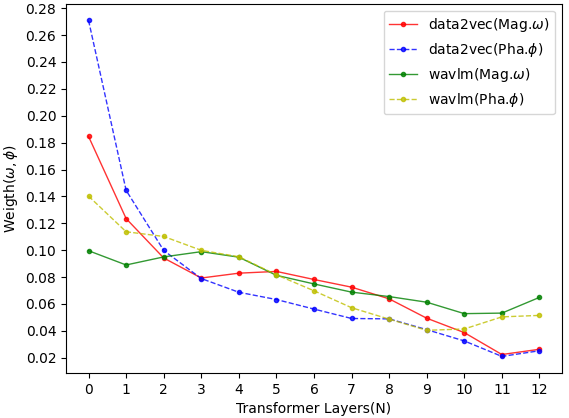}}
		\centerline{(b) Weighting analysis}\medskip
        \label{fig:f4_b}
	\end{minipage}
	\caption{The Matplotlib visualization analysis. (a) is 300 random samples from the VoiceBank+DEMAND test set were selected to compare the PESQ scores of BSP-MPNet with two baseline methods. (b) is An analysis of the weights corresponding to each Transformer layer in different SSL models.}
	\label{fig:f4}
\end{figure}

We also performed a visualization analysis, as shown in Fig. \ref{fig:f4}. (a) shows that our model outperforms others in the PESQ metric. (b) shows that for the SE task, the early Transformer layers in SSL are crucial, containing mainly acoustic information, while the later layers focus more on semantic information. This analysis is significant for applying SSL in downstream tasks and addressing information redundancy.
		
\section{CONCLUSION}	
This paper proposes the BSP-MPNet model and validates its effectiveness through experiments. Although the model outperforms other baselines, the original SSL models have a large number of parameters. 
Therefore, future work will combine the feature separation approach of FS-SSL with distillation techniques to further improve the model's efficiency.





\bibliographystyle{references/IEEEbib}
\bibliography{references/icme2025references}

\end{document}